\documentclass{aastex6}

\fullcollaborationName{The Friends of AASTeX Collaboration}

\begin{document}


\title{ Secular Orbit Evolution in Systems with a Strong External Perturber - A Simple and Accurate Model }


\author{Eduardo Andrade-Ines\altaffilmark{1}}
\affil{Institute de M\'ecanique C\'eleste et des Calcul des \'Eph\'em\'erides - Observatoire de Paris \\ 77 Avenue Denfert Rochereau \\ 75014 Paris, France}
\author{Siegfried Eggl\altaffilmark{2}}
\affil{Jet Propulsion Laboratory, California Institute of Technology \\
 4800 Oak Grove Drive \\ 91109 Pasadena, CA, USA}





\altaffiltext{1}{eandrade.ines@gmail.com}
\altaffiltext{2}{siegfried.eggl@jpl.nasa.gov}

\begin{abstract}


We present a semi-analytical correction to the seminal solution for the secular motion of a planet's orbit under gravitational influence of an external perturber derived by \citet{heppenheimer-1978}.
A comparison between analytical predictions and numerical simulations allows us to determine corrective factors for the secular frequency and forced eccentricity in the co-planar restricted three-body problem. 
The correction is given in the form of a polynomial function of the system's parameters that can be applied to first-order forced eccentricity and secular frequency estimates. The resulting secular equations are simple, straight forward to use and improve the fidelity of Heppenheimer’s solution well beyond higher-order models. The quality and convergence of the corrected secular equations are tested for a wide range of parameters and limits of its applicability are given.

\end{abstract}




\section{Introduction} 
\label{sec:intro}
Understanding the gravitational three-body problem is one of the fundamental challenges in celestial mechanics.
More than three centuries of study have produced a wide variety of methods to tackle the dynamics of three gravitating point masses \citep{marchal-1990}.
Although general analytical solutions do exist, they tend to be unwieldy due to the complexity of the problem \citep[e.g.][]{sundman1907recherches}. Thus, simplified  
approaches tailored to specific needs are still popular. In order to describe the long-term evolution of hierarchical three body systems,
for instance, fast-evolving variables can be eliminated from the equations of motion either through averaging or via canonical transformations of the corresponding
Hamiltonian system \citep{ferraz2007canonical}. The resulting ``secular equations'' are less cumbersome to deal with and permit simple 
analytical solutions to the dynamical evolution of the system.
%
In his seminal work on planet formation in binary stars, \citet{heppenheimer-1978} described the orbital evolution 
of a satellite influenced by an external perturber, e.g. 
a planet orbiting one component of a binary star {  considering all bodies to be point masses}\footnote{That is, neglecting the dynamical effects due to the shapes of the bodies.}. 
Assuming the planet started out on a circular orbit in the same 
orbital plane as the massive bodies, its eccentricity ($e_1$) and 
longitude of pericenter ($\varpi_1$) evolve as 
\begin{equation}
e_1(t)=\epsilon \big[4-4 \cos{(gt)}\big]^{1/4}, \qquad \varpi_1(t)= \frac{g}{2}\;t, 
\label{eq:hepp0}
\end{equation}
where $\epsilon$ and $g$ are the so-called \textit{forced eccentricity} and \textit{secular frequency}, respectively.
Both $\epsilon$ and $g$ are constants that depend on the system parameters as shown in Section \ref{sec:am}.
As in Heppenheimer's approximation the planet and the perturber only exchange angular momentum, specific orbital energies, semimajor axes and periods
stay constant for all bodies. Hence, Equations  (\ref{eq:hepp0}) completely describe the secular evolution of the system\footnote{This is assuming
that the planet does not influence the orbit of the perturber and that both the planet's and the perturber's angular momentum vectors are parallel.}.
The non-linear and singular aspects of the 
three-body problem, however, cause simplified solutions such as the one above to sometimes fail to capture the actual evolution of the system \citep{andrade2016secular}.

{  Today, direct numerical integration of the equations of motion is a viable option. 
While purely numerical methods have their own shortcomings,
they provide relatively accurate solutions on short to intermediate timescales \citep[e.g.][]{eggl2010introduction}. Yet, for large scale planet formation simulations
in particular, the time requirements to individually propagate each particle through numerical integrations are still prohibitive \citep{leiva2013mama,thebault2011against}.   
 A natural way to tackle such issues is to apply more complete analytic models, but those often contain higher-order expansions leading to a
substantial increase in complexity \citep{georgakarakos-2003,andrade2016secular}. }
Alternatively, one can try to modify lower order theories so as to better fit the behavior of the actual system under investigation \citep{giuppone-2011}. 

In this work we take the latter approach and derive an empirical correction to the classical solution of 
\citet{heppenheimer-1978} for the forced eccentricity and secular frequency.
Our corrections to Equations (\ref{eq:hepp0}) presented in this work improve the fidelity of Heppenheimer's solution substantially while still retaining its simple and elegant form.
The rest of the article is structured as follows.
In Section \ref{sec:am} we present a more detailed derivation of Heppenheimer's equations and discuss recent developments and extensions to the first-order secular approach.
Section \ref{sec:num_det} contains information on how we derive the corrections to the forced eccentricity and secular frequency from the integration 
of the exact equations of motion. 
Results for the corrections stemming from least squares fits for both the secular frequency and the forced eccentricity 
are presented in Section \ref{sec:corr} and in Section \ref{sec:app} we discuss some applications of the new model. 
Concluding remarks are provided in Section \ref{sec:conclusion}.

\section{Analytical Secular Orbit Evolution Models}
\label{sec:am}

\begin{figure}
\centering
\includegraphics[width=0.4\textwidth]{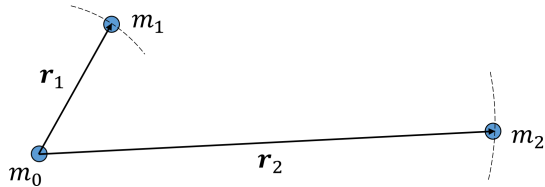}
\caption{Representation of the system consisted of a central body of mass $m_0$ being orbited by a satellite of (negligible) mass $m_1$. Another body of mass $m_2$ revolves around the primary so that its orbit never crosses the path of the satellite.}
\label{fig:def_vec}
\end{figure}

Let us consider a system consisting of a main body with mass $m_0$, a planet with mass $m_1$ and a secondary body with mass $m_2$ in a reference frame centered on $m_0$. Let $\vec r_i$ be the position vector of the $i$-th body with respect to $m_0$ (see Figure \ref{fig:def_vec}). We assume, furthermore, that all bodies move in the same plane. If 
$m_1 \ll m_0 \sim m_2$ and $|{\vec r_1}| < |\vec r_2|$ for all time and neglecting gravitational effects due to $m_1$, 
we arrive at the coplanar restricted three-body problem of S-type \citep{dvorak-1986}.
The orbit of $m_2$ around $m_0$ will be regarded as a fixed Keplerian ellipse, whereas the orbit of the planet evolves with time.
This section is dedicated to discussing previous analytical results that model the secular behavior 
of the host-planet-perturber configuration. Models by \cite{heppenheimer-1978}, \citet{marchal-1990} and \citet{andrade2016secular} have been selected for this purpose, as they represent
various degrees of complexity and fidelity with respect to the numerical reference solutions. 
The here presented analytic models will serve as a benchmark for the empirical correction derived in Section \ref{sec:corr}.  

\subsection{Heppenheimer (1978)}

\begin{figure}
\centering
\includegraphics[width=0.5\textwidth]{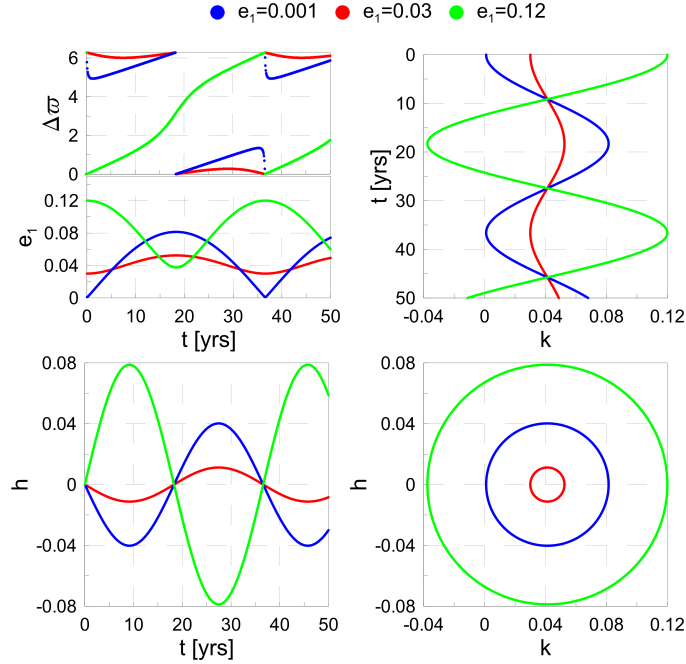}
\caption{Secular orbits obtained from Heppenheimer's model for a system with parameters $\mu=1$, $\alpha=0.1$ and $e_2 = 0.3$, 
integrated from orbits initially at the pericenter ($\Delta\varpi=0$) for three 
different initial planetary eccentricities: $e_1=0.001$ (blue), $e_1 = 0.03$ (red) and $e_1=0.12$ (green). 
Note that for all three systems $\epsilon_{H} = 0.041$. All circles in the $(k,h)$ plane (bottom right graph) are centered on ($\epsilon_H$, $0$).
Each of the three orbits evolves along their respective circle, and completes one revolution in $36.5$ years corresponding to a secular 
frequency of $g_H=0.172\; rad/yr$ regardless of their radius.}
\label{ex-secorb}
\end{figure}

Assuming there are no significant mean motion resonances, the long-term evolution of the planets's orbit ($m_1$) will be dominated by the secular interaction with
the perturber ($m_2$). The corresponding disturbing function can be expanded in terms of the semimajor axis ratio by means of Legendre polynomials ($P_i$). 
While this kind of expansion limits our model to hierarchical systems, it allows us to obtain finite expressions 
for arbitrary eccentricities \citep{kaula-1962,laskar2010explicit}. By limiting the expansion in Legendre polynomials to $P_3$ (quadrupole problem), 
truncating the perturbation to low values of the eccentricity of the planet and performing a first-order averaging, \citet{heppenheimer-1978} obtained the following secular disturbing 
function in orbital elements
\begin{equation}
{\cal R}_{H} = \frac{{\cal G} m_2}{(1-e_2^2)^{3/2}}\frac{a_1^2}{a_2^3}\left[ \frac{1}{4} + \frac{3}{8}e_1^2 - \frac{15}{16}\frac{a_1}{a_2}\frac{e_1 e_2}{(1-e_2^2)}\cos(\Delta\varpi) \right],
\label{hepp05}
\end{equation}
\noindent where ${\cal G}$ is the gravitational constant, $\Delta\varpi = \varpi_1-\varpi_2$, $a_i$, $e_i$ and $\varpi_i$ are the semimajor axis, the eccentricity and the longitude of the pericenter of the $i$-th body, respectively. 
Neglecting all constant terms as they will not change the system's equations of motion, Equation (\ref{hepp05}) can be expressed in a simpler form

\begin{equation}
{\cal R}_{H} = n_1 a_1^2 g_{H} \left[ \frac{e_1^2}{2} - \epsilon_{H} e_1 \cos\Delta \varpi \right],
\label{hepp1}
\end{equation}

\noindent where 

\begin{equation}
g_{H} = \frac{3}{4}\mu\alpha^3 \frac{n_1}{(1-e_2^2)^{3/2}}
\label{hepp2}
\end{equation}

\noindent is the secular frequency and

\begin{equation}
\epsilon_{H} = \frac{5}{4} \alpha \frac{e_2}{(1 - e_2^2)}
\label{hepp3}
\end{equation}

\noindent is the forced eccentricity, with $n_1=\sqrt{{\cal G}m_0/a_1^3}$ being the mean motion of the planet, $\alpha = a_1/a_2$ and $\mu = m_2/m_0$. Introducing the variables

\begin{eqnarray}
\begin{array}{rcl}
k &= e_1 \cos \Delta\varpi,\\
h &= e_1 \sin \Delta\varpi, 
\end{array}
\label{hepp4}
\end{eqnarray}
\noindent the Lagrange-Laplace planetary equations up to order $\mathcal{O}(e_1^2)$ that determine the satellite's orbit evolution read \citep{brouwer1961}:

\begin{equation}
\begin{array}{rcl}\vspace{0.2cm}
\displaystyle\frac{d h}{dt} = & \displaystyle\frac{1}{n_1 a_1^2}\displaystyle\frac{\partial {\cal R}}{\partial k} &= g_{H}\left(k-\epsilon_{H}\right),\\
 \displaystyle\frac{d k}{dt} = &- \displaystyle\frac{1}{n_1 a_1^2}\displaystyle\frac{\partial {\cal R}}{\partial h} &= - g_{H}h.
\end{array}
 \label{hepp5}
\end{equation}
The system of Equations (\ref{hepp5}) admits the analytical solution
\begin{equation}
\begin{array}{rl}
k =& e_p \cos(g_{H}t + \phi) + \epsilon_{H},\\
h =& e_p \sin(g_{H}t + \phi), 
\end{array}
\label{kt}
\end{equation}

\noindent where $e_p$ and $\phi$ are constants of integration, obtained from the initial conditions $e_{1}(0)$ and $\varpi_{1}(0)$ as 

\begin{equation}
e_p = \sqrt{(k_0-\epsilon_{H})^2 + h_0^2},
\label{ep}
\end{equation}
\begin{equation}
\cos \phi = \frac{k_0-\epsilon_{H}}{e_p} {\rm , \ } \sin \phi = \frac{h_0}{e_p}.
\label{phi}
\end{equation}
Here, $k_0 = e_{1}(0) \cos{\Delta\varpi(0)}$ and $h_0 = e_{1}(0) \sin{\Delta\varpi(0)}$, with $\Delta\varpi(0)=\varpi_1(0)-\varpi_2$. 
Figure \ref{ex-secorb} shows an example of how the time evolution of the secular variables $k$ and $h$ translates into the elliptic osculating elements $e_1$ and $\Delta\varpi$. 
Three different sets of initial conditions were chosen for the planet's  orbit, $e_1\in \{0.001,\; 0.03,\; 0.12\}$, sharing otherwise similar system parameters $\alpha = 0.1$, $\mu=1$ 
and $e_2 = 0.3$. The evolution of the elliptic elements $e_1$ and $\Delta\varpi$ is qualitatively different for each set of initial conditions. Yet,
the very same dynamical system can be described by simple circles in the $(k,h)$ plane. While the circles have different radii depending
on the initial conditions, they are all centered on  $(k,h)=(\epsilon_H,0)$.
This behavior is due to the form of Equations (\ref{kt}) that define circles in the $(k,h)$ plane.
The planet's orbit evolves along those circles with the frequency $g_{H}$. 
Additionally, we can see from Equations  (\ref{hepp2}) and (\ref{hepp3}) that $g_{H}$ and $\epsilon_{H}$ are functions of the 
parameters of the problem $n_1$, $\alpha=a_1/a_2$, $\mu=m_2/m_0$ and $e_2$ only. 
Those are constants of motion in the secular restricted three-body problem. 
Hence, for a fixed set of parameters, the secular orbits will always have the same forced eccentricity and the same secular frequency, 
no matter the initial values of $\varpi_1$ and $e_1$.\footnote{This statement holds for models up to order ${\mathcal O}(e_1^2)$.}   
As the planet’s eccentricity increases, higher-order terms become more important and the secular frequency may change from the 
predicted value. However, this is only expected to happen for $e_1>0.2$ \citep{andrade2016secular}. 
If the planet's orbit was initially circular, i.e. $h(0)=k(0)=0$, so that $e_p=\epsilon$ and $\phi=\pi$, then Equations  (\ref{kt})
reduce to Equations (\ref{eq:hepp0}) presented in the introduction.

\subsection{Marchal (1990)}

Using the Von Zeipel averaging method on the three-body disturbing function, \citet{marchal-1990} obtained a second-order\footnote{in $\mu$ and $\alpha$, see also \citet{georgakarakos-2003,georgakarakos-2005}} 
solution for the secular orbit evolution of an S-type hierarchical triple system. In the following
we briefly discuss Marchal's model and the resulting equations.
Still considering the planar problem, the disturbing function of Marchal's model is given by
\begin{equation}
{\cal R}_M = Q_1 + Q_2 + Q_3
\label{geor1}
\end{equation}
\noindent where 
\begin{equation}
Q_1 = \displaystyle\frac{1}{4} {\cal G}\displaystyle\frac{m_0m_2}{m_0+m_1}\displaystyle\frac{a_1^2}{a_2^3} \left(1+\displaystyle\frac{3}{2} e_1^2 \right) \displaystyle\frac{1}{(1-e_2^2)^{3/2}},
\label{geor2}
\end{equation}
\begin{equation}
\begin{array}{rl}
\vspace{0.4cm}
Q_2 =& -\displaystyle\frac{15}{16} {\cal G}\displaystyle\frac{m_0m_2(m_0-m_1)}{(m_0+m_1)^2} \displaystyle\frac{a_1^3}{a_2^4} \\
\vspace{0.4cm}
& \times \ e_1\left( 1+\displaystyle\frac{3}{4} e_1^2 \right) \displaystyle\frac{e_2}{(1-e_2^2)^{5/2}} \cos \Delta\varpi
\end{array}
\label{geor3}
\end{equation}
\noindent originate in the Legendre polynomial expansion up to $P_3$ and
\begin{equation}
\begin{array}{rl}
\vspace{0.4cm}
Q_3 = & \displaystyle\frac{15}{64} {\cal G}\displaystyle\frac{m_0m_2^2}{(m_0+m_1)^{3/2}(m_0+m_1+m_2)^{1/2}} \\
\vspace{0.4cm}
& \times \ e_1^2 (1-e_1^2)^{1/2}\displaystyle\frac{\left[5(3 + 2e_2^2) + 3e_2^2 \cos(2\Delta\varpi) \right]}{(1-e_2^2)^3}
\end{array}
\label{geor4}
\end{equation}
\noindent arises from the Von Zeipel's averaging method as a second-order term. 
Assuming now the restricted problem ($m_1/m_0 \to 0$) and neglecting terms of order $\mathcal{O}(e_1^3)$, Equation (\ref{geor1}) becomes 
\begin{equation}
{\cal R}_M = n_1 a_1^2 g_{M}\left[ \frac{e_1^2}{2}  - \epsilon_{M} e_1 \cos\Delta \varpi \right],
\label{geor5}
\end{equation}
\noindent where
\begin{equation}
g_{M} = g_{H} \left[1 + \frac{25}{8}\frac{\mu}{\sqrt{1+\mu}}\alpha^{3/2} \frac{3+2e_2^2}{(1-e_2^2)^{3/2}} \right],
\label{gsge1}
\end{equation}
\begin{equation}
\epsilon_{M} = \epsilon_{H} \left[1 + \frac{25}{8}\frac{\mu}{\sqrt{1+\mu}}\alpha^{3/2} \frac{3+2e_2^2}{(1-e_2^2)^{3/2}} \right]^{-1}.
\label{efge1}
\end{equation}
The disturbing function ${\cal R}_M$ in Equation (\ref{geor5}) has
a form similar to ${\cal R}_H$ in Equation (\ref{hepp1}) where $g_{M}$ and $\epsilon_{M}$ are the quantities corresponding
to $g_{H}$ and $\epsilon_{H}$, respectively. 
Therefore, the same general form of solutions given by Equations  (\ref{kt}) can be expected to work. Note that, although of similar form, the expressions for the secular 
frequency and forced eccentricity are not identical in the two models.
Equations   (\ref{gsge1}) and (\ref{efge1}) relate Marchal's secular quantities with those in Heppenheimer's model.
Only if $\mu\ll 1$, (\emph{i.e.} $m_2\ll m_0$), both models share the same analytic expressions.

\subsection{Andrade-Ines et al. (2016)}
Utilizing Hori's averaging theory \citep{hori-lie_series1966,ferraz2007canonical}, 
\citet{andrade2016secular} recently constructed a second-order\footnote{in $\alpha^2 \cdot \mu$} model that, 
too, provided more accurate results than Heppenheimer's solution.
Relying on an expansion of the disturbing function in powers of the 
semimajor axis ratio and eccentricities, however, the model is complex and a complete analytical form is not available at this time. 
The model developed by \citet{andrade2016secular} works as follows.
The second-order Hamiltonian (${\cal H}_{A}$) and the secular frequencies $g_{A}$ are calculated from analytical expressions presented in 
\citet{andrade2016secular} 
using tabulated numerical coefficients of the disturbing function. The forced eccentricity ($\epsilon_{A}$) is a root of the Equation
\begin{equation}
\frac{\partial {\cal H}_{A}}{\partial e_1} \Big|_{e_1=\epsilon_{A},\Delta\varpi=0} = 0.
\label{modogs3}
\end{equation}
Equation (\ref{modogs3}) can be solved for $\epsilon_{A}$ numerically by applying, for instance, 
the geometric method presented in \citet{michtchenko2004-secular}. 
Finally, the secular orbits for the second-order Hamiltonian can be obtained from Equations  (\ref{kt}), 
replacing $g_{H}$ and $\epsilon_{H}$ by the numerical values obtained for $g_{A}$ and $\epsilon_{A}$, respectively.

%

\section{Extracting secular solutions from direct N-body simulations}
\label{sec:num_det}

So far, we have seen that the secular problem we are investigating can be reduced to one degree of freedom associated with the constant frequency $g$.
As $g$ only depends on fixed parameters of the problem ($\alpha$, $\mu$ and $e_2$), it can be calculated in advance. 
Returning to Figure \ref{ex-secorb} we see that the secular evolution of the planet's orbit in the 
the $(k,h)$ plane can be described with circles centered at $(\epsilon_{},0)$. 
The radius of such a circle is equal to $e_p$ which, similar to $\phi$, depends exclusively on the system's initial conditions. 
In particular for $e_p = 0$ we have a \emph{stationary secular solution}, with $e_1(t) = \epsilon$ and $\Delta\varpi(t) = 0$. 
In other words, the planet's orbit ($a_1$, $e_1$,$\varpi_1$) does not change its shape and keeps its prior orientation with respect to the perturber's orbit,
($\dot a_1=\dot e_1=\dot \varpi_1=0$).
In the complete\footnote{non-secular, i.e. not averaged over the fast angles} planar problem, however, we have three degrees of freedom, 
associated with the following three frequencies: the mean motions of the planet ($n_1$) and the perturber ($n_2$) and the secular frequency $g$. 
In this case, the stationary secular solution 
corresponds to a quasi-periodic orbit in osculating elements with frequencies $n_1$ and $n_2$
and zero amplitude in all quantities associated with the secular frequency $g$. 
Therefore, the orbits and the initial conditions of the complete problem can be fundamentally different 
from those of the secular (averaged) problem, which makes the comparison between the two models a non-trivial task.
Methods based on the frequency analysis \citep{laskar1990-freqmap,michtchenko2002-chaosmap} 
of numerically integrated orbits have 
been used successfully by many authors to determine such quasi-periodic solutions \citep{noyelles2008-decomp,couetdic-2010-decomp,andrade2016secular}. 
In this work, we chose to apply the method described in \citet{andrade2016secular} that was developed for the case of stationary secular solutions in the restricted three-body problem.
For a fixed set of parameters, this method finds the quasi-periodic secular solution after successive iterations of a digital filter 
that eliminates the secular component of the time series of the variable $\xi=e_1 \textrm{e}^{{\textrm i}\Delta\varpi}$. Once the initial conditions of the 
quasi-periodic orbit are determined, the secular frequency $g_{E}$ is obtained via harmonic decomposition, 
while the forced eccentricity is calculated as
\begin{equation}
\epsilon_{E} = \langle e_1(t) \cos{\Delta\varpi(t)} \rangle_T, 
\label{eFnum}
\end{equation}

\noindent 
where $T$ is the total time of integration. 
Unless the secular period $P_E=2\pi/g_{E}$ is known exactly, this process yields the true values for $\epsilon_{E}$ 
only if the total time of integration tends to infinity. 
However, good approximations can be found for sufficiently long\footnote{at least several secular periods} times of integration. 
Fortunately, as the secular component was already eliminated from the quasi-periodic solution, 
the only periodic components still present are the fast frequencies $n_1$ and $n_2$. 
Therefore, by considering a time of integration of that is much longer than the Keplerian period of the outer body, 
the error in the determination of $\epsilon_E$ is going to be small as well. In our calculations, 
we considered a time of integration of $T = 12\pi/g_H$, that is 6 times the secular period of the system. 
Throughout this work we shall consider the solutions for $\epsilon_{E}$ and $g_{E}$ obtained from direct N-body simulations to be the most
accurate ones. They will serve as calibration values for the results originating from analytic and semi-analytic methods.
Though fast and accurate, the exclusively numerical methodology to calculate secular quantities has its limitations as well. 
For instance, if the orbits are chaotic or in a mean motion resonance (MMR) the method will not converge \citep{andrade2016secular}. 
Similarly, care has to be taken when comparing initial conditions used for the same orbit in both, the complete problem and the models. 
A given set of osculating initial conditions of the complete problem is not the same as the averaged initial conditions of the secular models, 
specially for the case when there are strong short-period perturbations involved \citep{andrade2016secular}. 
In this work, we consider as initial conditions of the secular problem $\big(k(0),h(0)\big)$ the average of the same variables 
$\big(k(t),h(t)\big)$ over one orbital period of the outer body calculated with the complete problem.

\subsection{Results of the Comparison}

\begin{figure}
\centering
\includegraphics[width=0.5\textwidth]{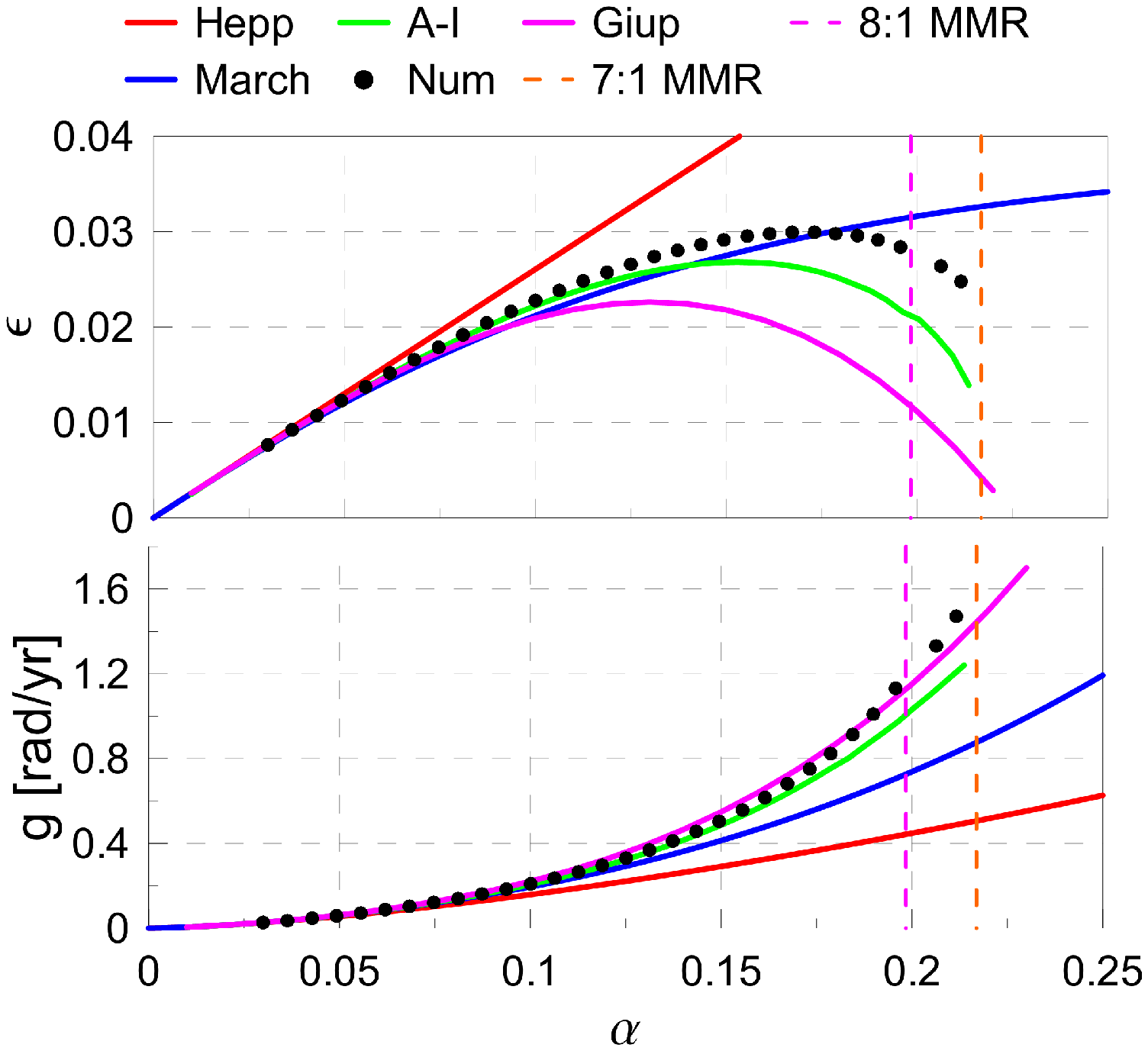}
\caption{This figure contains forced eccentricity (top) and secular frequency (bottom) estimates as a function of the semimajor axis ratio for systems sharing the parameters $m_0= 1\;M_\odot$, $m_1 = 10^{-5}\;M_\odot$, $a_2 = 1\;$au, $\mu=1$ and $e_2=0.2$.
Only the satellite's semimajor axis $a_1$ is varied.
Results are shown for the following models: \citet{heppenheimer-1978} (red curve), \citet{marchal-1990} (blue curve), \citet{andrade2016secular} (green curve) and {  \citet{giuppone-2011} (magenta curve)} models.
The numerically determined solution can be considered the most accurate one (black dots). 
The vertical dashed lines represent the nominal positions of the 7:1 (orange) and 8:1 (magenta) mean motion resonances. 
{  The numerical solution is omitted for $\alpha < 0.025$ 
as this is a region where the external perturbations tend to zero and all models start to coincide.} }
\label{ex-m1e02}
\end{figure}

In order to check the fidelity of the methods presented in the previous sections, we calculated the stationary secular solutions for 
several systems using each of the previously mentioned models.
Figure \ref{ex-m1e02} contains a comparison among all models, where the fully numerically computed, secular results serve as a reference. 
The model by \citet{heppenheimer-1978} is referred to as \emph{Hepp}, \citet{marchal-1990}'s model as \emph{March} and the one by
\citet{andrade2016secular} as \emph{A-I}. 
{  In the same figure we include the results given by the empirical correction derived by \cite{giuppone-2011} (referred as \emph{Giup}) 
designed for the $\gamma$ Cephei system.} The same abbreviations remain valid throughout the remainder of this work.
One can see that, for small values of the semimajor axis ratio $\alpha$, all the models coincide with the reference solution. 
For large $\alpha$ values, however, none of the models provides accurate representations of the system dynamics.
It becomes evident from Figure \ref{ex-m1e02}, though, that
more complex models of higher order tend to work better for larger values of $\alpha$. {  Although \emph{Giup} is an empirical approximation to a second-order analytical model 
it provides a poor fit to the forced eccentricity for the particular case presented in Figure \ref{ex-m1e02}. 
This is understandable, as the approximation was derived to study the $\gamma$ Cephei system and shall be only applicable to systems 
sharing similar parameters.} 
The gap in the curve of the reference solution at $\alpha \approx 0.2$ is due to the 7:1 mean motion resonance (MMR) between the satellite and the perturber. 
As discussed in the previous section, crossing a MMR impedes the convergence of the iterative method used to construct the reference solution. Such gaps are going to appear whenever MMRs begin to be significant for the dynamical evolution.

%

\section{Introduction of Empirical Corrections}
\label{sec:corr}
A comparison between Equations  (\ref{hepp1}) and (\ref{geor5}) shows that, by limiting expansions to low eccentricities of the satellite,
we can write the Hamiltonian in a quadratic form permitting a general solution akin to Equations  (\ref{kt}).
Since \textit{Hepp} and \textit{March} models are not identical, differences in the actual values of the secular frequency and the forced eccentricity are to be expected.
A glance at Equations  (\ref{gsge1}) and (\ref{efge1}) reveals 
that the higher-order terms can be interpreted as corrections to the first-order solutions for the forced eccentricity $\epsilon_H$ and the secular frequency $g_H$.  
The second-order solution can take a very complex form, though, and worse, it may still not describe 
the secular motion for all stable orbits in the relevant parameter-space with the desired accuracy \citep{andrade2016secular}. 
Figure \ref{ex-m1e02} contains some of the systems where the second-order model does not match the reference solution at all, especially for large semimajor
axis ratios. Those systems may, perhaps, be properly described by third- or higher-order solutions, which, unfortunately, 
further increase in complexity.

In order to circumvent this issue, one can take an alternative approach by introducing \emph{empirical} corrections. Those are 
coefficients added to a simple model, which are constructed from the reference solution through a fitting procedure. 
This was successfully done by \citet{giuppone-2011}, where the authors presented corrected values of $g$ and $\epsilon$.
Those, however, were only valid for the particular case of the $\gamma$-Cephei system. In this work, we will present a general expression for these quantities, valid for a much wider range of parameters.
Let us define the relative difference (error) between the reference solution and Heppenheimer's model in the secular frequency $g_{H}$ and the forced eccentricity $\epsilon_{H}$, respectively, as
\begin{equation}
 \delta_{g}(\alpha,e_2,\mu) = 1 - \frac{g_{E}}{g_{H}},
\label{empcorr1}
\end{equation}
\begin{equation}
\delta_{\epsilon}(\alpha,e_2,\mu) = 1 - \frac{\epsilon_{E}}{\epsilon_{H}},
\label{empcorr2}
\end{equation}
\noindent where $g_{E}$ and $\epsilon_{E}$ are obtained numerically. 
Those errors are evaluated for several mass ratios of the binary $\mu = (0.1, 0.2, 0.5, 1, 2, 5, 10)$, 
given a set of binary eccentricities $e_2 = (0.1,0.2,0.3,0.4,0.5,0.6)$ and semimajor axis ratios $\alpha \in [0.01,0.4]$. 
The sampling interval of the latter was varied between $0.01$ to $0.05$ depending on the system, such that there were at least $25$ different values of $
\alpha$ for each pair of parameters $(\mu,e_2)$. {  For all simulations, we fixed the mass of the central body at $m_0 = 1M_\odot$, and the semimajor axis of the perturber at $a_2 = 1 \textrm{au}$.} We then applied the method of least squares to minimize the error functions $\delta_{g}$ and $\delta_{\epsilon}$, assuming that the correction functions have a polynomial dependence on the parameters of the problem $\mu = m_2/m_0$, $\alpha = a_1/a_2$ and $e_2$, in the form of 
\begin{equation}
\delta_{g} = \sum_{i=1}^{N_g} A_i \alpha^{p_i} e_2^{q_i} \mu^{l_i},
\label{empcorr3}
\end{equation}
\begin{equation}
\delta_{\epsilon} = \sum_{i=1}^{N_e} A^\prime_i \alpha^{p_i^\prime} e_2^{q_i^\prime} \mu^{l_i^\prime}.
\label{empcorr4}
\end{equation}
Naturally, one can expect the larger the number of terms in the fit, the smaller the total error. 
However, the higher the order of the fitting polynomials, the faster the fit deviates from the nominal solution outside of the fitting region.
Furthermore, high-order polynomials quickly become unwieldy themselves, defeating the purpose of simplifying the secular model. Limiting our expressions to less than 20 terms, a series of different functions $\delta_{\epsilon}$ and $\delta_{g}$ was tested and we applied an iterative method 
to determine the optimum number of terms and the order $(p_i,q_i,l_i)$ of the polynomial functions {  that resulted in the smallest error for the fitted quantities}. The results are the following expressions:
\begin{equation}
\begin{array}{rl}
\vspace{0.2cm}
\delta_{g} = \alpha^{3/2} & [-4.6274 \mu^{1/2}-4.0190 \mu+0.25041 \mu^{2} \\
\vspace{0.2cm}
& -3.41 e_2^2 \mu^{1/2}+11.09 e_2^2 \mu-0.9823 e_2^2 \mu^{2} \\
\vspace{0.2cm}
& -20.13 e_2^4 \mu^{1/2}-85.49 e_2^4 \mu+4.996 e_2^4 \mu^{2}] \\
\vspace{0.2cm}
+\alpha^{9/2} & [123.67 \mu^{1/2}-799.20 \mu -201.49 \mu^{2} \\
\vspace{0.2cm}
&+180  e_2^2 \mu^{1/2} -5555  e_2^2 \mu-617.7 e_2^2 \mu^{2} \\
\vspace{0.2cm}
&+2.671\times 10^{4}  e_2^4 \mu^{1/2}-1.0229\times 10^5 e_2^4 \mu\\
&-23076 e_2^4 \mu^{2}], \\
\end{array}
\label{corr_freq_min}
\end{equation}
\begin{equation}
\begin{array}{rl}
\vspace{0.2cm}
\delta_{\epsilon} = \alpha^{3/2} & [29.494 e_2 \mu^{1/2} + 9.220 e_2 \mu -99.85 e_2^2 \mu^{1/2}\\
\vspace{0.2cm}
&-31.50 e_2^2 \mu + 124.60 e_2^3 \mu^{1/2} + 35.69 e_2^3 \mu] \\
\vspace{0.2cm}
+\alpha^{9/2} & [1073.0 e_2 \mu^{1/2}+4280 e_2 \mu-1609.8 e_2 \mu^{2} \\
\vspace{0.2cm}
&-4161 e_2^2 \mu^{1/2} -2.978\times 10^4 e_2^2 \mu+6429 e_2^2 \mu^{2}\\
\vspace{0.2cm}
&+1.82\times 10^3 e_2^3 \mu^{1/2}+7.449\times 10^4 e_2^3 \mu \\
& -8681 e_2^3 \mu^{2}]. \\
\end{array}
\label{corr_ecc_min}
\end{equation}

The fit deduced uncertainties of the coefficients in the above Equations are tabulated in Appendix 
\ref{app_coeff}. {  Note that the above formulae do not represent a series expansion in the parameters $\alpha$, $\mu$ and $e_2$. 
They represent the best multivariate polynomial fit to numerical simulations. As such, the coefficients do not need to 
fulfil convergence criteria. In fact, 
including additional terms in the fit functions would drastically change the numerical coefficients presented.} 
Given the above expressions for $\delta_g$ and $\delta_\epsilon$ the corrections to the secular frequency and the forced eccentricity read: 
\begin{equation}
g_{C} = g_{H}[1-\delta_{g}],
\label{gcorrected}
\end{equation}

\begin{equation}
\epsilon_{C} = \epsilon_{H}[1-\delta_{\epsilon}].
\label{ecorrected}
\end{equation}
Applying this correction is straight-forward. 
The secular orbit evolution can be obtained simply by replacing $g_{H}$ and $\epsilon_{H}$ by $g_{C}$ and $\epsilon_{C}$, respectively, in Equations (\ref{kt}).
The resulting secular quantities are valid for small satellite  eccentricities and for the 
range of parameters $0.1 \leq e_2 \leq 0.6$ and $0.1 \leq m_2/m_0 \leq 10$, for all stable and non-resonant $\alpha$.

\subsection{Quality of the Correction}

\begin{figure*}
\centering
\includegraphics[width=0.9\textwidth]{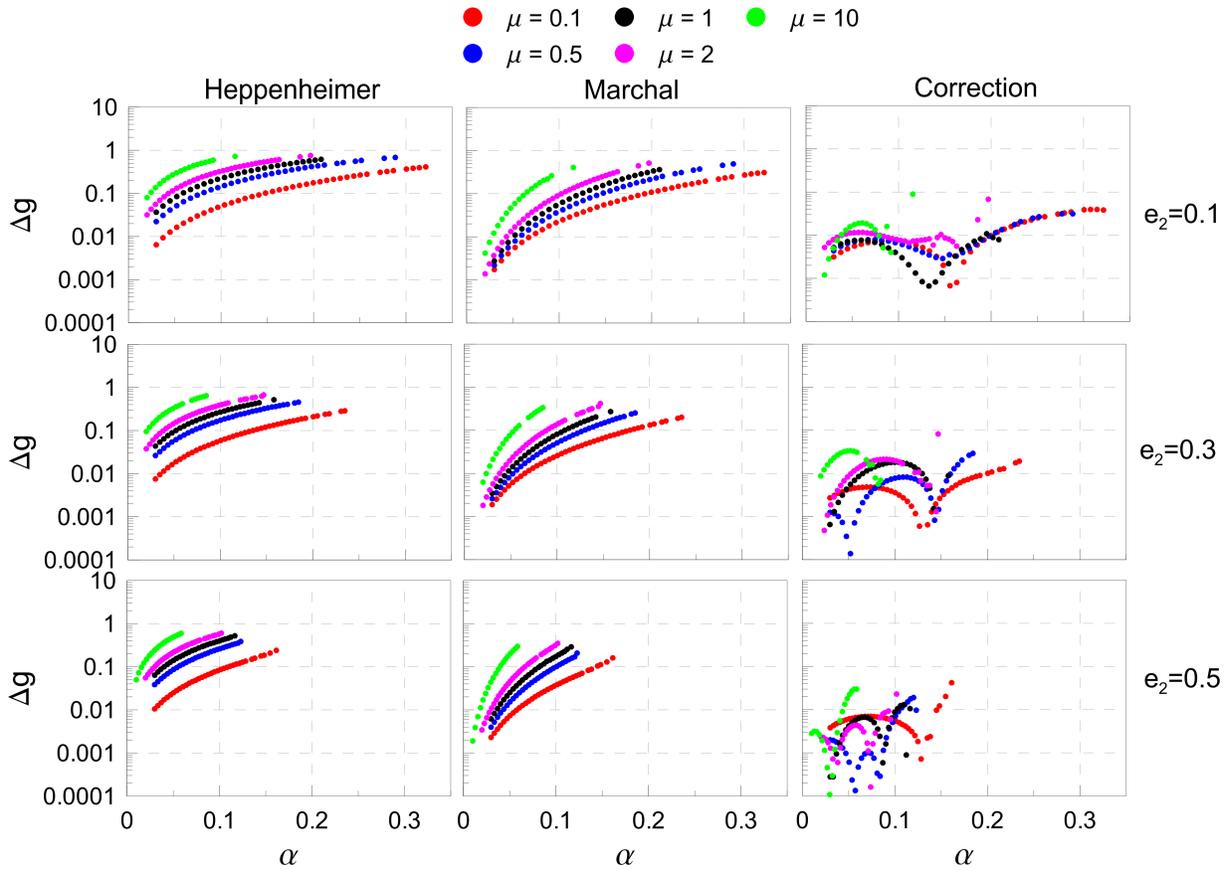}
\caption{ Relative error of the secular frequency $g$ calculated using \textit{Hepp} (left column), \textit{March} (middle column) and \textit{Corr} (right column)
with respect to the reference solution as a function of the semimajor axis ratio ($\alpha$). 
The three rows represent different perturber eccentricity values ($e_2$) and the color code shows results for various mass ratios of the binary ($\mu$). 
{  Similar to Figure \ref{ex-m1e02}, values for $\alpha < 0.025$ were not calculated as the perturbations become negligible and all methods 
start to provide accurate approximations to the planetary orbit.}}
\label{error_gs}
\end{figure*}

\begin{figure*}
\centering
\includegraphics[width=0.9\textwidth]{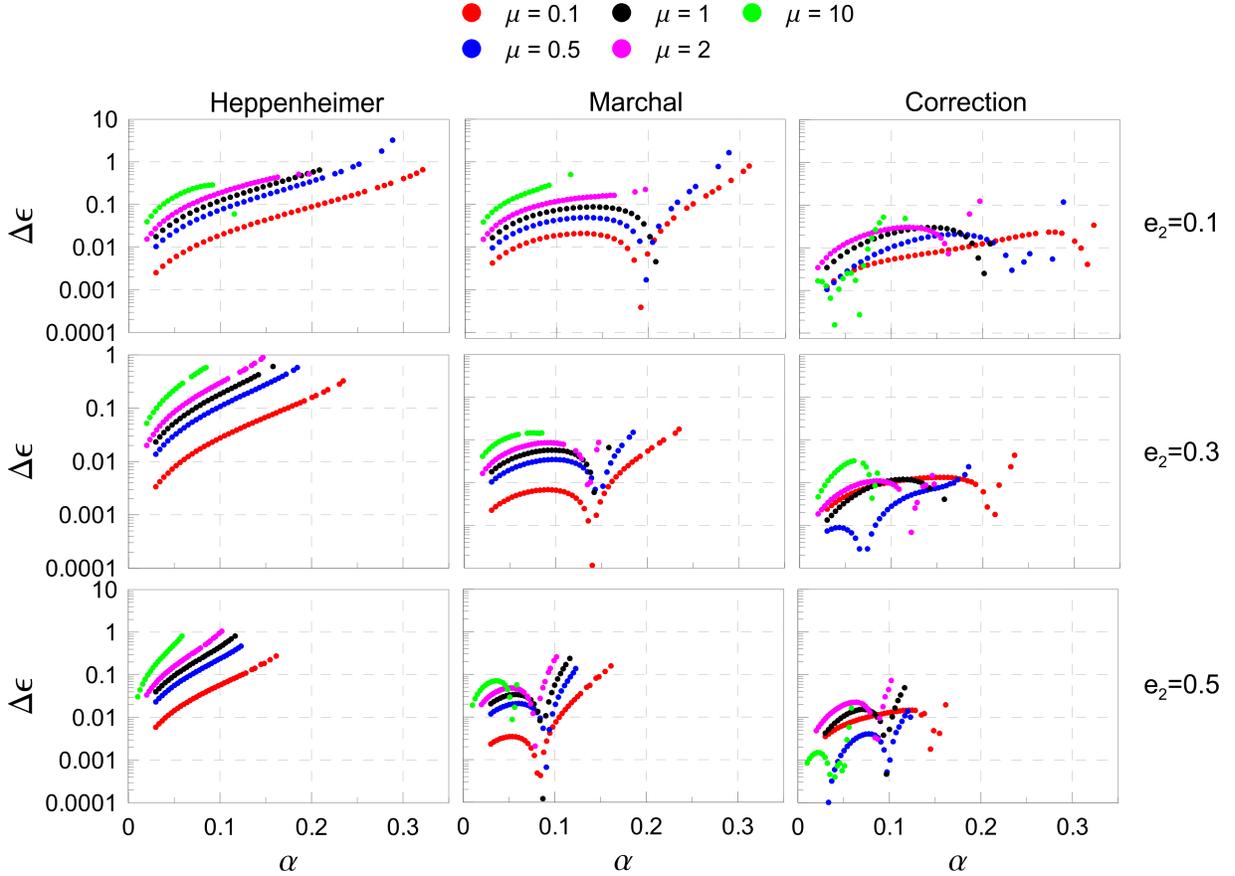}
\caption{Same as Figure \ref{error_gs} but for the forced eccentricity $\epsilon$.}
\label{error_ef}
\end{figure*}

The corrections presented in Equations  (\ref{corr_freq_min}) and (\ref{corr_ecc_min}) are finite polynomial 
approximations and, therefore, are still bound to deviate from the reference solution. 
To assess the quality of the correction 
we compare results obtained from the analytical models by Heppenheimer and Marchal with those achieved using Heppenheimer plus our correction and 
gauge them on the reference solution.\footnote{We have omitted the \textit{A-I} model in this comparison for computational reasons, as the evaluation takes somewhat longer than for the other methods.}
In order to quantify the deviation of the respective approach from the reference solution  
we define the standard relative error as

\begin{equation}
 \Delta x_{y} = \frac{|x_{y} - x_{ref}|}{x_{ref}},
\end{equation}

\noindent where $x=(g,\epsilon)$ are the secular parameters, the index $ref$ stands for the reference solution obtained from the N-body simulations and $y=(Hepp,March,Corr)$ represent the model choice. 
Figures \ref{error_gs}  and \ref{error_ef} show the relative errors in the secular quantities as a function of the semimajor axis ratio $\alpha$.

Both, Figures \ref{error_gs} and \ref{error_ef} underline that the results obtained from the corrected model (\textit{Corr}) 
are excellent. The errors of the corrected secular frequencies as well as forced eccentricities 
are much smaller than those resulting from any other method. 
For the majority of the parameter-space ($\alpha,e_2,\mu$), the relative error is smaller than 5\%. 
The points exhibiting the largest errors are all located close to the stability limits in very localized regions of the ($\alpha,e_2,\mu$) space. 
The non-optimum behavior there is caused by the presence of significant mean motion resonances that the numerical method itself is not able to resolve 
(see Figure \ref{ex-m1e02}). 
The left panels in Figures \ref{error_gs} and \ref{error_ef} show that the error for \textit{Hepp} increases with larger values of $\alpha$, $e_2$ and $\mu$, which is to be expected. 
The panels in the middle column of both Figures \ref{error_gs} and \ref{error_ef} show that Marchal's model leads to
much smaller errors compared to Heppenheimer's, especially for small values of $e_2$, $\mu$. 
Regarding the forced eccentricity, \textit{March} presents a peculiar non-monotonous behavior: 
the error begin increasing, then decrease and finally, for some cases, cross the origin while increasing again. 
While this suggests that \textit{March} is applicable to a wider range of parameters, the same is not true for the secular frequency (Figure \ref{error_gs}).
%
%
%

\section{Applicability and discussion of the empirically corrected secular solution}
\label{sec:app}
\begin{figure*}[!ht]
\centering
\includegraphics[width=0.9\textwidth]{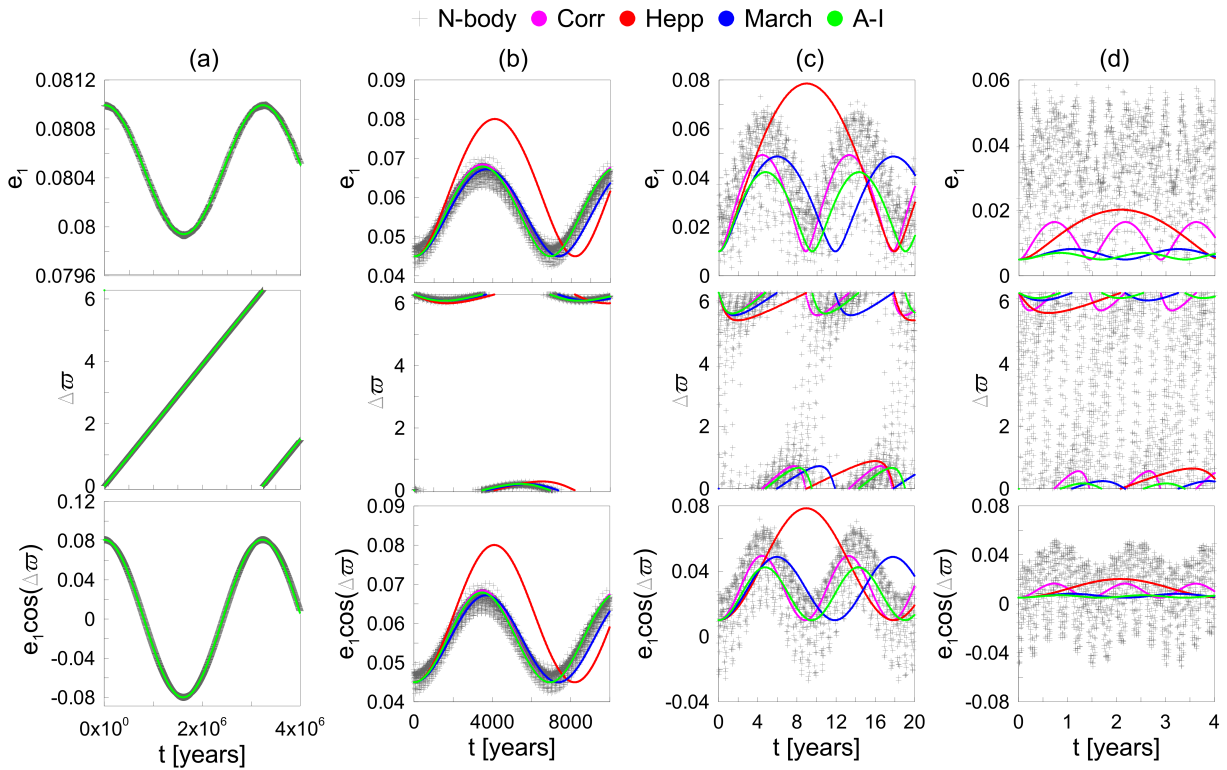}
\caption{Examples of the orbits obtained from the integration of the different models for the systems presented in Table \ref{tab:example} compared to the integration of the exact equations of motion.}
\label{ex_int_orb}
\end{figure*}

In order to demonstrate possible applications and limits of applicability of the corrected secular solution presented in this work,
we have selected four different S-type three-body configurations. Two systems correspond to actually detected exoplanets 
orbiting a single star with a giant planet perturber, and one component of a binary star. The other two systems are fictitious, selected
to showcase the limits of the various secular theories.
Their physical and orbital parameters are presented in Table \ref{tab:example}. {   We also include  in Table \ref{tab:example} the values obtained by \textit{Hepp} and \textit{Corr} for the forced eccentricity and the secular frequency for each of the systems as well as the the stability limits.} 
Figure \ref{ex_int_orb} compares the orbit evolution of the satellite in the systems given in Table \ref{tab:example} 
as predicted by the four secular models (\textit{Hepp, March, A-I, Corr}) to
the results of a direct N-body integration including non-secular and short period variations. 

\begin{table}
\centering
\caption{Initial parameters of the examples {  and secular parameters calculated with \emph{Hepp} and \emph{Corr} models for the forced eccentricity and secular frequency}.} 
\label{tab:example}
\begin{tabular}{lllll}
\hline
       & HD 41004Bb$^{a}$ {\bf(a)} & $\gamma$ Cephei Ab$^{b}$ {\bf(b)} & {\bf(c)} & {\bf(d)} \\ \hline
$m_0(M_\odot)$ & 0.42 & 1.4 & 1 & 1 \\
$m_1(M_\odot)$ & $1.743 \times 10^{-2}$ & $1.765 \times 10^{-3}$ & $10^{-4}$ & $10^{-4}$ \\
$m_2(M_\odot)$ & 0.7 & 0.41 & 1 & 10 \\
$ a_1 (au)$ & 0.0177 & 2.05 & 0.17 & 0.1 \\
$ a_2 (au)$ & 20.0 & 20.2 & 1 & 1 \\
$ e_1 $ & 0.081 & 0.05 & 0.01 & 0.01 \\
$ e_2 $ & 0.4 & 0.41 & {  0.2} & {  0.1} \\
$\epsilon_H$ 	&	$5.27 \times 10^4$	&	0.063	&	0.044	&	0.0123	\\
$\epsilon_C$ 	&	$5.27 \times 10^4$	&	0.057	&	0.030	&	0.0103	\\
$g_H (rad/yr)$ 	&	$1.95\times 10^{-6}$	&	$7.66 \times 10^{-4}$	&	0.351	&	1.46	\\
$g_C (rad/yr)$ 	&	$1.95\times 10^{-6}$	&	$9.01 \times 10^{-4}$	&	0.709	&	3.98	\\
\hline
\end{tabular} \\
\tablenotetext{a}{\citet{santos2002,zucker2004,roell-et-al-2012}}
\tablenotetext{b}{\citet{neuhauser2007,endl2011,reffert2011}}
\tablecomments{We assumed for simplicity $M_1=M_2=\varpi_1=\varpi_2=0$ and we considered masses to be maximal ($m_1 \sin i = m_1$).}
\end{table}
The HD 41004 Bb system, example {\bf(a)}, is characterized by a very hierarchical structure ($\alpha \ll 1$) with little short period activity.
All models were able to properly describe its secular dynamics. 
In contrast, the planet orbiting $\gamma$ Cephei A, presented as example {\bf(b)}, 
experiences much stronger perturbations, since the system hosts a seconday star relatively close to the primary.
This causes \textit{Hepp} to no longer provide accurate estimates for the secular frequency and forced eccentricity. 

If one considers even stronger perturbations, such as in example {\bf(c)}, for instance,
we see that both \textit{Hepp} and \textit{March} start to diverge from the N-body integration. Note that short period terms that are not considered in secular dynamics start to play an important role even far from mean motion resonances \citep{georgakarakos-2003,georgakarakos-2005}.
Finally, in example {\bf(d)}, only \textit{Corr} is able to capture the secular dynamics of the system, even when short period terms are dominating the direct N-body integration result.
{  
Due to the high amplitude short-period oscillations present in both the eccentricity $e_1$ and the secular angle $\Delta \varpi$, 
the secular variable $k=e_1\cos(\Delta\varpi)$ reaches negative values. 
In both examples  {\bf(c)} and {\bf(d)} the free eccentricity $e_p$ is smaller than the forced one $\epsilon$ 
which leads to a secular oscillation of $\Delta \varpi$ around 0 (see Figure \ref{ex-secorb}). However, in the complete problem, the secular angle alternates between circulation and oscillation around $\Delta\varpi=0$. 
This particular characteristic may give the impression that none of the models is capable of reproducing the the N-body integration results for $e_1$.
Nevertheless, we note that the secular variables $(k,h)$ can still be accurately described by \textit{Corr}.}
High amplitude short-period oscillations are also an indication that higher-order theories should be applied when constructing analytical models to properly describe secular dynamics \citep{andrade2016secular}.

%
\section{Conclusions}
\label{sec:conclusion}
Simple and accurate models for the long-term orbit evolution of small satellites under the influence of an external perturber are highly desirable, as they 
find multiple applications in various fields of astrophysics. 
Although several analytic models exist that describe moderately perturbed systems sufficiently well,  
tend to become unwieldy and imprecise in the case of strong perturbations. In this work we have mapped out an escape-route from this dilemma 
by providing a simple means to calculate the secular dynamics of heavily perturbed three body systems
based on empirical corrections of the simple model by \citet{heppenheimer-1978}. 
Those corrections were obtained through a fitting procedure bringing secular frequency ($g$) and forced eccentricity ($\epsilon$) predictions 
closer to more precise results obtained from direct N-body integrations.  
The resulting Equations (\ref{gcorrected}) and (\ref{ecorrected}) are extremely simple and straight-forward to apply.
We have shown that our empirical correction can describe all stable secular orbits of planets 
in hierarchical three body configurations of `S-type' near perfectly for a wide range of initial conditions, i.e. all semimajor axis ratios $\alpha< 0.4$,
perturber eccentricities $e_1 < 0.6$, and perturber mass ratios $0.1 \leq \mu \leq 10$.
Applying our predictions to several exoplanetary systems, we confirm an excellent agreement between our simple secular model and direct
N-body calculation results. 

\acknowledgments
This research has in part been funded by CNPq Project 204873/2014-2 and by 
the Jet Propulsion Laboratory through the California Institute of Technology postdoctoral fellowship program, under a contract with the National Aeronautics and Space Administration. 
The authors would, furthermore, like to acknowledge the support of the IMCCE, Observatoire de Paris, France. The authors would like to thank the anonymous referee who helped to improve this manuscript.







\appendix

\section{Coefficients and Errors of the Correction}
\label{app_coeff}
In Table \ref{tab:errors} the coefficients used in the empirical correction factors for the planet's secular frequency ($g$, Equation (\ref{empcorr3}))  and its forced eccentricity ($\epsilon$, Equation (\ref{empcorr4})) are presented. 
The fit-derived uncertainties for the parameters are also shown. 
\begin{table*}[!ht]
\centering
\caption{Numerical coefficients of the empirical corrections for the secular frequency $A_i$ (Equation (\ref{empcorr3}))and for the forced eccentricity $A_i^\prime$ (Equation (\ref{empcorr4})) and their errors $\sigma A_i$ and $\sigma A_i^\prime$, respectively.} 
\label{tab:errors}
\begin{tabular}{llllrllllrl}
\hline
$i $ & $ p_i $ & $ q_i $ & $ l_i $ & $ A_i $ & $ \sigma A_i $ & $ p_i^\prime $ & $ q_i^\prime $ & $ l_i^\prime $ & $ A_i^\prime $ & $ \sigma A_i^\prime$ \\ \hline
$1 $ & $ 1.5 $ & $ 0 $ & $ 0.5 $ & $ -4.6274 $ & $ 0.0058 $ & $ 1.5 $ & $ 1 $ & $ 0.5 $ & $ 29.494 $ & $ 0.056 $\\
$2 $ & $ 1.5 $ & $ 0 $ & $ 1 $ & $ -4.0190 $ & $ 0.0052 $ & $ 1.5 $ & $ 1 $ & $ 1 $ & $ 9.220 $ & $ 0.032 $\\
$3 $ & $ 1.5 $ & $ 0 $ & $ 2 $ & $ 0.25041 $ & $ 0.00039 $ & $ 1.5 $ & $ 2 $ & $ 0.5 $ & $ -99.85 $ & $ 0.30$\\
$4 $ & $ 1.5 $ & $ 2 $ & $ 0.5 $ & $ -3.41 $ & $ 0.11 $ & $ 1.5 $ & $ 2 $ & $ 1 $ & $ -31.50 $ & $ 0.17 $\\
$5 $ & $ 1.5 $ & $ 2 $ & $ 1 $ & $ 11.09 $ & $ 0.10 $ & $ 1.5 $ & $ 3 $ & $ 0.5 $ & $ 124.60 $ & $ 0.36 $\\
$6 $ & $ 1.5 $ & $ 2 $ & $ 2 $ & $ -0.9823 $ & $ 0.0077 $ & $ 1.5 $ & $ 3 $ & $ 1 $ & $ 35.69 $ & $ 0.20 $\\
$7 $ & $ 1.5 $ & $ 4 $ & $ 0.5 $ & $ -20.13 $ & $ 0.31 $ & $ 4.5 $ & $ 1 $ & $ 0.5 $ & $ 1073.0 $ & $ 7.5 $\\
$8 $ & $ 1.5 $ & $ 4 $ & $ 1 $ & $ -85.49 $ & $ 0.29 $ & $ 4.5 $ & $ 1 $ & $ 1 $ & $ 4280 $ & $ 14 $\\
$9 $ & $ 1.5 $ & $ 4 $ & $ 2 $ & $ 4.996 $ & $ 0.023 $ & $ 4.5 $ & $ 1 $ & $ 2 $ & $ -1609.8 $ & $ 3.5 $\\
$10 $ & $ 4.5 $ & $ 0 $ & $ 0.5 $ & $ 123.67 $ & $ 0.48 $ & $ 4.5 $ & $ 2 $ & $ 0.5 $ & $ -4161 $ & $ 67 $\\
$11 $ & $ 4.5 $ & $ 0 $ & $ 1 $ & $ -799.20 $ & $ 0.88 $ & $ 4.5 $ & $ 2 $ & $ 1 $ & $ -2.978 \times 10^{4} $ & $ 1.2\times 10^{2} $\\
$12 $ & $ 4.5 $ & $ 0 $ & $ 2 $ & $ -201.49 $ & $ 0.22 $ & $ 4.5 $ & $ 2 $ & $ 2 $ & $ 6429 $ & $ 28 $\\
$13 $ & $ 4.5 $ & $ 2 $ & $ 0.5 $ & $ 180 $ & $ 23 $ & $ 4.5 $ & $ 3 $ & $ 0.5 $ & $ 1.82\times 10^{3} $ & $ 1.3\times 10^{2} $\\
$14 $ & $ 4.5 $ & $ 2 $ & $ 1 $ & $ -5555 $ & $ 40 $ & $ 4.5 $ & $ 3 $ & $ 1 $ & $ 7.449\times 10^{4} $ & $ 2.3\times 10^{2} $\\
$15 $ & $ 4.5 $ & $ 2 $ & $ 2 $ & $ -617.7 $ & $ 8.8 $ & $ 4.5 $ & $ 3 $ & $ 2 $ & $ -8681 $ & $ 54 $\\
$16 $ & $ 4.5 $ & $ 4 $ & $ 0.5 $ & $ 2.671\times 10^{4} $ & $ 15 $ & $ - $ & $ - $ & $ - $ & $ - $ & $ -$\\
$17 $ & $ 4.5 $ & $ 4 $ & $ 1 $ & $ -1.0229\times 10^{5} $ & $ 2.6\times 10^{2} $ & $ - $ & $ - $ & $ - $ & $ - $ & $ - $\\
$18 $ & $ 4.5 $ & $ 4 $ & $ 2 $ & $ -23076 $ & $ 54 $ & $ - $ & $ - $ & $ - $ & $ - $ & $ - $\\
\hline
\end{tabular}
\end{table*}

\section{Notation}
Table \ref{tab:decimal} contains a list of the variables used in this work together with a brief description. 
\begin{table*}[!ht]
\centering
\caption{Description of the notation in this work.} 
\label{tab:decimal}
\begin{tabular}{rl}
\hline
Variable & Description\\
\hline
$m_i$ & Mass (central body: $i=0$, planet: $i=1$, secondary body: $i=2$)\\
$\mu$ & Mass ratio $m_2/m_0$ \\
$a_i$ & Semimajor axis (planet: $i=1$, secondary body: $i=2$)     \\
$\alpha$ & Semimajor axis ratio $a_1/a_2$ \\
$e_i$ & Eccentricity (planet: $i=1$, secondary body: $i=2$)     \\
$\varpi_i$ & Longitude of the pericenter (planet: $i=1$, secondary body: $i=2$) \\
$\cal G$ & Gravitational constant \\
$P_i$ & Legendre polynomials \\
$n_i$ & Mean motion (planet: $i=1$, secondary body: $i=2$) \\
$k,\; h$ & Coordinates of the planet's eccentricity vector \\
$t$ & Time \\
$k_0,\; h_0$ & Initial values of the coordinates of the eccentricity vector\\
$g_{y}$ & Secular frequency (\textit{Hepp}: $y=H$, \textit{March}: $y=M$, \textit{A-I}: $y=A$, \textit{Corr}: $y=C$) \\
$\epsilon_{y}$ & Forced eccentricity (\textit{Hepp}:  $y=H$, \textit{March}: $y=M$, \textit{A-I}: $y=A$, \textit{Corr}: $y=C$) \\
$e_p$ & Proper eccentricity \\
$\phi$ & Initial phase \\
$\delta_x$ & Correction term (secular frequency: $x=g$, forced eccentricity: $x=\epsilon$)\\
$A_i,\; A^\prime_i$ & Numerical correction coefficients \\
$p_i,q_i,l_i,\; p^\prime_i,q^\prime_i,l^\prime_i$ & Exponents of $\alpha$, $e_2$ and $\mu$ in the correction  \\
$\Delta x_{y}$ & Error (secular frequency: $x=g$, forced eccentricity: $x=\epsilon$) \\ 
& for each model (\textit{Hepp}: $y=H$, \textit{March}: $y=M$, \textit{A-I}: $y=A$, \textit{Corr}: $y=C$) \\
\hline
\end{tabular}
\end{table*}

\bibliographystyle{hapj} 
\bibliography{diss} 


\listofchanges

\end{document}